\documentclass[11pt,a4paper]{article}
\pdfoutput=1

\usepackage{jheppubedit}
\usepackage{enumitem}
\usepackage{amsmath}
\usepackage{amsfonts}
\usepackage{graphicx}
\usepackage{caption}

\title{The long string at the stretched horizon and the entropy of large non-extremal black holes}

\author[a,b]{Thomas G. Mertens,}
\author[b]{Henri Verschelde}
\author[c,d,e]{and Valentin I. Zakharov}
\affiliation[a]{Joseph Henry Laboratories, Princeton University, Princeton, NJ 08544, USA,} 
\affiliation[b]{Ghent University, Department of Physics and Astronomy\\
Krijgslaan, 281-S9, 9000 Gent, Belgium}
\affiliation[c]{ITEP, B. Cheremushkinskaya 25, Moscow, 117218 Russia,}
\affiliation[d]{Moscow Inst Phys \& Technol, Dolgoprudny, Moscow Region, 141700 Russia ,}
\affiliation[e]{School of Biomedicine, Far Eastern Federal University, Sukhanova str 8, 
Vladivostok 690950 Russia}

\emailAdd{tmertens@princeton.edu}
\emailAdd{henri.verschelde@ugent.be}
\emailAdd{vzakharov@itep.ru}
\abstract{We discuss how long strings can arise at the stretched horizon and how they can account for the Bekenstein-Hawking entropy. We use the thermal scalar field theory to derive the asymptotic density of states and corresponding stress tensor of a microcanonical long string gas in Rindler space. We show that the equality of the Hagedorn and Hawking temperatures gives rise to the tree-level entropy of large black holes in accordance with the Bekenstein-Hawking-Wald formula.}

\begin{document}

\maketitle

\section{Introduction}
The statistical origin of the Bekenstein-Hawking entropy of a black hole is still in many respects an unsolved problem. By comparing the classical black hole laws and the second law of thermodynamics, it follows that a black hole has a classical entropy given by 
\begin{equation}
S_{BH} = \frac{A}{4G},
\end{equation}
where $A$ is the area of the event horizon and $G$ Newton's gravitational constant. A simple understanding of how this classical result arises from counting underlying quantum microstates is still lacking. If one throws in particles described by local quantum field theory, a fiducial observer close to the horizon will see a hot gas of ever-increasing local temperature as the matter sinks in towards the horizon, contributing a divergent thermal entropy. As was noted by 't Hooft \cite{'tHooft:1984re}, this means that the horizon can store an infinite amount of information and hence this problem is equivalent with the information problem for black holes. To describe the physics at the horizon correctly, one needs a consistent theory of quantum gravity, even if the curvature at the horizon of large black holes is vanishingly small.  \\

\noindent An appealing solution to the divergent entropy problem was proposed by Susskind \cite{Susskind:1993ws}, who argued that if one throws in strings instead of particles, the strings will undergo a percolation transition to one long string close to the horizon. Since the long string has a maximal temperature, it will hover above the horizon at a distance of the order of the string length $\ell_s = \sqrt{\alpha'}$ where the local temperature is of the order of the Hagedorn temperature.\footnote{An alternative interpretation for this hovering string was given in \cite{Hewitt:1993he}.}\\
The long string is therefore a realization of the stretched membrane known from the membrane paradigm approach to black hole physics \cite{Thorne:1986iy}. A revolution in our understanding of black holes was launched by the introduction of D-brane techniques in string theory. For extremal black holes made from D-branes, one could count BPS microstates using supersymmetry protection against interactions and agreement with the classical Bekenstein-Hawking formula was found \cite{Strominger:1996sh}. For near-extremal and non-extremal black holes, the state counting is not protected by supersymmetry and the situation is less clear. Nevertheless, there exist some corners of parameter space where apparently interaction effects can still be neglected. Interestingly, in these cases the entropy can be modeled as coming from a single long closed string with a rescaled tension and effective central charge $c_{\text{eff}}=6$. \\

\noindent Very recently, the study of strings in the near-horizon region of a black hole received renewed interest \cite{Giveon:2012kp}\cite{Giveon:2013ica}\cite{Giveon:2014hfa}\cite{Giveon:2013hsa}\cite{Martinec:2014gka}\cite{Halyo:2015ffa}\cite{Halyo:2015oja}\cite{Giveon:2015cma}\cite{Giribet:2015kca}\cite{Ben-Israel:2015mda}\cite{Silverstein:2014yza}\cite{Dodelson:2015toa}\cite{Dodelson:2015uoa}. \\

\noindent In this paper, we present a simple picture of black hole entropy taking into account interactions in a sort of mean field way. We build up the black hole by throwing in shells of closed strings one after another. We start from a nucleus black hole with mass $M_0$, which is large enough and sufficiently non-extremal so that it can be described close to the horizon by a Rindler metric. Then we throw in a shell of closed strings with mass $\delta M$ from infinity which are treated as non-interacting with each other but propagating in the background of the nucleus black hole. The mass of the shell is chosen to be sufficiently small so that backreaction and the self-interaction within the shell is negligible. As the shell sinks towards the horizon, it percolates into a long string at $\rho \sim \ell_s$ from the horizon. Once the shell has settled, we throw in another shell which now propagates in the background of a black hole with mass $M_0+\delta M$. We sum over shells until the final mass is reached. We show that the accumulated entropy of the shells accounts precisely for the increase of entropy of the nucleus black hole as dictated by the Bekenstein-Hawking area law. The crucial result we use is that the density of states of a long string in a Rindler background is given by a Hagedorn distribution with Hagedorn temperature equal to the Unruh temperature as shown in \cite{Mertens:2013zya}. \\

\noindent Section \ref{longflat} contains the results about the flat space long string gas from a thermal scalar perspective. Section \ref{longrind} generalizes these results to Rindler space. We pay special attention to the stress tensor of these long strings. Our main result is contained in section \ref{mainr} where we demonstrate that the Bekenstein-Hawking entropy can be obtained from these long equilibrated strings close to the horizon. Some conclusions are presented in section \ref{concl} and appendix \ref{rotat} contains the generalization of this story to rotating black holes.

\section{The long string gas in flat space}
\label{longflat}
We consider a gas of non-interacting strings in flat space at temperature $1/\beta$ \cite{Mitchell:1987hr}\cite{Mitchell:1987th}. The free energy of the gas is given by
\begin{equation}
\label{fre}
\beta F = \sum_n \rho(E_n)\ln\left(1-e^{-\beta E_n}\right) = -\sum_{w=1}^{\infty}\sum_n \rho(E_n) \frac{e^{-w\beta E_n}}{w},
\end{equation}
where $w$ can be interpreted as a winding number of the strings around the thermal circle and $\rho(E_n)$ is the single string density of states. This result is for bosonic strings. For spacetime fermions, one has instead
\begin{equation}
\beta F = -\sum_n \rho_F(E_n)\ln\left(1+e^{-\beta E_n}\right) = -\sum_{w=1}^{\infty}\sum_n \rho_{F}(E_n) \frac{(-)^{w-1}e^{-w\beta E_n}}{w}.
\end{equation}
If the theory is spacetime supersymmetric, each bosonic state has a fermionic partner and the total free energy becomes:
\begin{equation}
\beta F = -\sum_{w=1}^{\infty}\sum_n \rho(E_n) \frac{e^{-w\beta E_n}}{w}( 1 +(-)^{w-1})
\end{equation}
which implies even-$w$ sectors are absent. In more detailed string computations at genus one in the modular strip domain, one can interpret this $w$-quantum number as the single torus wrapping number in the strip \cite{Mertens:2014cia}. The detailed expressions for the flat space type II superstring indeed exhibit an absence of even-$w$ sectors \cite{Alvarez:1986sj}. \\

\noindent The asymptotic single string density of states can be obtained from the free energy of the non-interacting string gas in the following way: since at high energy, we have a Hagedorn spectrum:
\begin{equation}
\rho(E_n) \sim e^{\beta_H E_n},
\end{equation} 
it follows from the $w=1$ term in (\ref{fre}) that the free energy will start to diverge around $\beta \approx \beta_H$. Note that higher winding modes are unimportant, they only kick in at $\beta \approx \beta_H/w$. Therefore, around $\beta_H$, we have 
\begin{equation}
\beta F \approx -\sum_n \rho(E_n) e^{-\beta E_n}.
\end{equation}
We can therefore use the free energy of the free string gas as a technical device to calculate asymptotic densities of single string states. It does not mean that we advocate the canonical ensemble for black hole thermodynamics. In this paper, we will use the microcanonical ensemble throughout with as input the asymptotic single string density of states as obtained from the canonical free energy. \\

\noindent In \cite{Polchinski:1985zf}, Polchinski showed that the free energy of the free string gas can be written as the torus partition function of the string on the thermal manifold. The string spectrum on the thermal manifold now contains winding modes around the thermal circle. In the tachyon channel we have for type II superstrings:
\begin{equation}
m^2 = \frac{4\pi^2n^2}{\beta^2} + \frac{w^2\beta^2}{4\pi^2\alpha'^2} - \frac{2}{\alpha'},
\end{equation}
and for Kaluza-Klein momentum $n=0$, we obtain massless states at
\begin{equation}
\beta_w = \frac{2\sqrt{2}\pi\sqrt{\alpha'}}{w}.
\end{equation}
The $w=1$ mode corresponds with the Hagedorn temperature:
\begin{equation}
\beta_H = 2 \sqrt{2}\pi\sqrt{\alpha'}
\end{equation}
and following \cite{Atick:1988si}\cite{Horowitz:1997jc}\cite{Barbon:2004dd}, we can interpret the Hagedorn transition as a condensation of the $w=1$ mode, also called the thermal scalar. This mode is a complex scalar living in one dimension less and is not projected out by GSO. \\

\noindent To derive the Hagedorn spectrum from the free energy, we calculate the dominating contribution of the thermal scalar to the free energy. From the thermal scalar action:
\begin{equation}
S[\phi] = \int d^{d-1}x \sqrt{G}\left[\nabla_i \phi \nabla^i \phi^* + M(\beta)^2 \phi\phi^*\right],
\end{equation}
with $M(\beta)^2 = \frac{1}{4\pi\alpha'^2}\left(\beta^2-\beta_H^2\right)$, we have:
\begin{align}
\beta F &\sim \text{Tr}\ln\left(-\nabla^2+M(\beta)^2\right) \\
&\sim - \int_{0}^{+\infty}\frac{d\tau}{\tau} \text{Tr}e^{-\left(-\nabla^2+M(\beta)^2\right)\tau} \\
&\sim - \int_{0}^{+\infty}\frac{d\tau}{\tau} \sum_{n} e^{-\lambda_n(\beta)\tau},
\end{align}
where $\lambda_n(\beta)$ are the eigenvalues of the thermal scalar. Assuming compact dimensions, the lowest eigenmode dominates close to the Hagedorn temperature:
\begin{equation}
\lambda_0(\beta) = M(\beta)^2 \sim \frac{\beta_H}{2\pi\alpha'^2}\left(\beta-\beta_H\right)
\end{equation}
and becomes a zero-mode. \\
Introducing the energy variable $E= \frac{\beta_H \tau}{2\pi\alpha'^2}$, the dominant contribution to the free energy around $\beta \approx \beta_H$ can be written as
\begin{equation}
\beta F \sim -\int_{0}^{+\infty} dE \rho(E) e^{-\beta E},
\end{equation}
with
\begin{equation}
\rho(E) = \frac{e^{\beta_H E}}{E}.
\end{equation}
If there are non-compact dimensions, this procedure also recovers the necessary prefactors \cite{Horowitz:1996nw}. \\

\noindent It can be shown that this thermal winding state also encodes the fact that highly excited strings behave as random walks in the ambient space \cite{Kruczenski:2005pj}\cite{theory}. \\

\noindent Let us now consider a long string gas in the microcanonical ensemble with energy $E$. In \cite{Horowitz:1997jc}\cite{Mertens:2014dia} it was shown that the average value of the Euclidean energy-momentum tensor at high energy is given by the energy-momentum tensor of the thermal scalar field theory, evaluated on the zero-mode:
\begin{equation}
\left\langle T^{\mu\nu}\right\rangle = T^{\mu\nu}_{th.sc.}(\phi_0).
\end{equation}
For flat space, the properly normalized zero-mode is non-normalizable and we find (in finite-volume regularization)
\begin{equation}
\phi_0 = \sqrt{\frac{E}{V}}\frac{\sqrt{\alpha'}}{2}.
\end{equation}
The spatial stress tensor vanishes: $T_{ij} = 0$. The long string gas in flat space is therefore pressureless. As we will see in the next section, it develops pressure in a gravitational field.

\section{The long string gas in Rindler space}
\label{longrind}
The thermal scalar action in Rindler space is \cite{Giveon:2013ica}\cite{Mertens:2013zya}
\begin{equation}
S = \int d^{d-1}x \sqrt{G}e^{-2\Phi}\left[G_{ij}\nabla^i \phi \nabla^j \phi^* + \frac{1}{4\pi^2\alpha'^2}\left(\beta^2G_{00}-\beta_H^2\right)\phi\phi^*\right],
\end{equation}
with 
\begin{equation}
\label{rindmetr}
ds^2 = a^2\rho^2 dt^2 + d\rho^2 + d\mathbf{x}_{\perp}^2.
\end{equation}
In \cite{Giveon:2013ica}\cite{Mertens:2013zya}, it was shown that for type II strings, the thermal scalar action gets no $\alpha'$ corrections for Rindler space. We argued that this is also true for heterotic strings (but not for the bosonic case) \cite{Mertens:2014saa}. \\
The equation for the eigenmodes, independent of $\mathbf{x}_{\perp}$, is:
\begin{equation}
\left[-\partial_\rho^2 - \frac{1}{\rho}\partial_\rho + \frac{1}{4\pi^2\alpha'^2}\left(\beta^2a^2\rho^2-\beta_H^2\right)\right]\phi_n(\rho) = \lambda_n \phi_n(\rho).
\end{equation}
For type II strings, the normalizable eigenmodes and corresponding eigenvalues are given by \cite{Mertens:2013zya}:
\begin{align}
\phi_n(\rho) &= \exp\left(-\frac{a\beta \rho^2}{4\pi\alpha'}\right)L_n\left(\frac{a\beta \rho^2}{2\pi \alpha'}\right), \\
\lambda_n &= \frac{1}{\pi\alpha'}\left(a\beta(1+2n)-2\pi\right),
\end{align}
where $L_n$ are Laguerre polynomials and $n= 0,1,2,\hdots$. \\
At $\beta = 2\pi/a$, the Rindler temperature, the lowest eigenmode becomes a (normalizable) zero-mode concentrated around $\rho\sim\sqrt{\alpha'}$:
\begin{equation}
\label{zerom}
\phi_0 = N \exp\left(-\frac{\rho^2}{2\alpha'}\right).
\end{equation}
It dominates the free energy:\footnote{We used $E = \frac{a\tau}{\pi\alpha'}$.}
\begin{align}
\beta F &\sim -\int_{0}^{+\infty}\frac{d\tau}{\tau}e^{-\frac{1}{\pi \alpha'}\left(a\beta-2\pi\right)\tau} \\
&= -\int_{0}^{+\infty}\frac{dE}{E}e^{\frac{2\pi E}{a}}e^{-\beta E},
\end{align}
so that, using the same techical device as in flat space, we obtain the asymptotic density of states in Rindler space as:
\begin{equation}
\label{dos}
\rho(E) = \frac{e^{\beta_R E}}{E}.
\end{equation}
So, the Hagedorn temperature in Rindler space is simply the Unruh temperature $\beta_R = 2\pi /a$! Note that the radial direction behaves as a compact direction.\\

\begin{figure}[h]
\centering
  \includegraphics[width=0.25\linewidth]{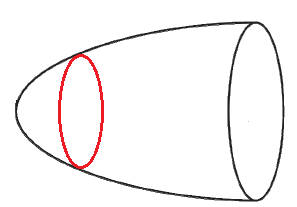}
  \caption{A cigar-shaped manifold allows winding strings to have arbitrarily small length.}
  \label{ciga}
\end{figure}

\noindent Some questions immediately come to mind. Why is the Hagedorn temperature so low and not of order $1/\ell_s$, and why does the thermal scalar not become tachyonic and condense? Let's first answer the second question. Naively, one expects tachyonic behavior close to the horizon because a winding string in Euclidean Rindler space can pinch off from the tip of the cigar (figure \ref{ciga}) so that the winding scalar becomes tachyonic.
However, in a gravitational field, the long string gas develops a pressure gradient, keeping the winding string away from the tip of the cigar. This phenomenon is analogous to the pressure gradient in the earth's atmosphere which keeps it from falling onto the earth surface. The origin of  the 
pressure however is not kinetic but entropic. As one can check \cite{Mertens:2014dia}, a microcanonical ensemble describing a long string gas with energy $E$ at infinity (which means $\rho = 1/a$ where the temperature equals $1/\beta$):\footnote{We follow the signature convention $(-++\hdots +)$.}
\begin{equation}
E =  -\int d^{d-1}x \sqrt{G}T^{0}_{0}
\end{equation}
is described by the thermal scalar zero-mode (\ref{zerom}) with
\begin{equation}
N = \sqrt{\frac{E}{A}}\sqrt{\frac{1}{a\alpha'}},
\end{equation}
where $A$ is the area of the horizon. For the energy-momentum tensor, one finds \cite{Mertens:2014dia}:
\begin{align}
\label{stressen}
\left\langle T^{0}_{0}(\mathbf{x})\right\rangle &= - 2N^2\left(\frac{2\rho^2}{\alpha'}-1\right)e^{-\rho^2/\alpha'}, \\
\left\langle T^{\rho\rho}(\mathbf{x})\right\rangle &= 2N^2e^{-\rho^2/\alpha'}, \\
\left\langle T^{ij}(\mathbf{x})\right\rangle &= \delta_{ij}2N^2\left[1- \frac{\rho^2}{\alpha'}\right]e^{-\rho^2/\alpha'},
\end{align}
whose spatial profiles are shown below (figure \ref{form}):
\begin{figure}[h]
\centering
\begin{minipage}{.3\textwidth}
  \centering
  \includegraphics[width=\linewidth]{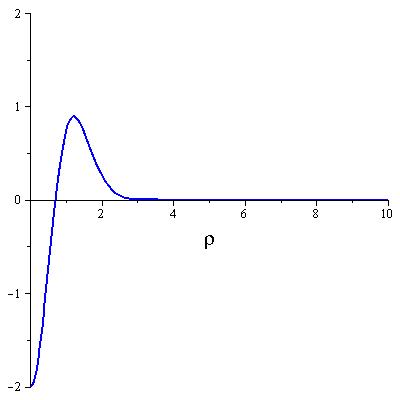}
  \caption*{(a)}
\end{minipage}
\begin{minipage}{.3\textwidth}
  \centering
  \includegraphics[width=\linewidth]{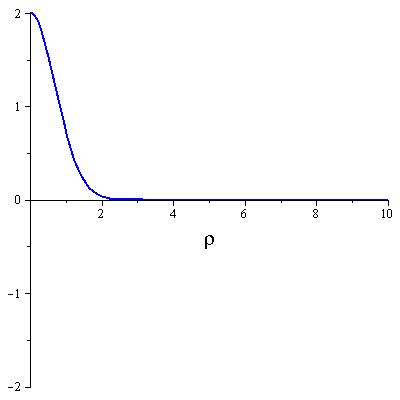}
  \caption*{(b)}
\end{minipage}
\begin{minipage}{.3\textwidth}
  \centering
  \includegraphics[width=\linewidth]{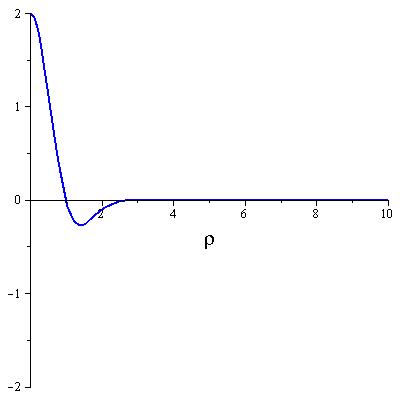}
  \caption*{(c)}
\end{minipage}
\caption{(a) Energy density $-\left\langle T^{0}_{0}(\mathbf{x})\right\rangle$ as a function of radial distance $\rho$ in units where $\alpha'=1$. (b) Radial pressure $\left\langle T^{\rho}_{\rho}(\mathbf{x})\right\rangle$. (c) Transverse pressure $\left\langle T^{i}_{i}(\mathbf{x})\right\rangle$.}
\label{form}
\end{figure}
The radial pressure $T^{\rho}_{\rho}$ keeps the long string gas from falling on the horizon. The random walk view of long strings provides us with another way of understanding this. The configurations of the long string in the microcanonical ensemble are the Feynman paths of a non-relativistic thermal scalar particle in an extremal attractive potential determined by the gravitational background \cite{theory} and the uncertainty principle keeps the particle far enough from the horizon so that its bound state energy is positive for $\beta > \beta_R$. Note that the radial pressure profile differs from the transverse pressure. This phase of matter is hence anisotropic, which is of course no real surprise for a 1d object.\\

\noindent The answer to the first question is now obvious since, because of (\ref{zerom}), the long string is living at $\rho \sim \sqrt{\alpha'}$ where there is a blueshift factor of order $\frac{1}{a\sqrt{\alpha'}}$. \\

\noindent When $\beta \neq 2\pi/a$, the spaces (\ref{rindmetr}) represent conical spaces. We make two remarks here. Firstly, such spaces are not saddle points of the path integral in canonical gravity. However, insisting on keeping fixed the horizon area during the variation, requires the introduction of a Lagrange multiplier which effectively introduces stress-energy at the origin \cite{Susskind:1994sm}\cite{Carlip:1993sa}. This causes the conical spaces to be valid saddle points after all when computing the thermodynamic entropy. \\
Secondly, we note that string theory on conical spaces is highly non-trivial: modular invariants are known only for $\mathbb{C}/\mathbb{Z}_N$ for integer $N$.\footnote{See however \cite{Mertens:2015adr}, where we construct modular invariant expressions for any conical deficit. These partition functions however ignore the surface contributions, a feature that is non-physical when considering the full thermal ensemble. The results we present here are based on the partition functions that do encode these surface terms.} The benefit of the above approach is that we are only interested in a single string mode for which we focus on the relevant second quantized field theory. Since the latter is well-defined for arbitary conical deficits, we do not run into trouble with this conundrum. \\

\noindent Finally we remark that also heterotic strings can be treated analogously. The thermal scalar wave equation contains discrete momentum as well, but in the end a zero-mode is again found at the Unruh temperature. This implies also heterotic strings have an equality between the Hagedorn and Hawking temperatures of the black hole. Bosonic strings on the other hand turn out to be a bit more subtle and we refer the interested reader to our previous work \cite{Mertens:2013zya}.

\section{Bekenstein-Hawking and Wald entropy of large black holes}
\label{mainr}
Consider a large Schwarzschild black hole of mass $M$ equiped with its thermal atmosphere. Let us now throw in a shell of mass $\delta M \ll M$ from infinity. We model the shell as a microcanonical closed string gas. We assume that the string coupling constant is non-zero but small so that as the shell sinks in towards the horizon, it equilibrates as a long random walking string at the stretched horizon. Since close to the horizon, the metric is Rindler, the long string has a Hagedorn density of states
\begin{equation}
\rho(\delta M_{\text{sh}}) \sim \frac{e^{2\pi\sqrt{\alpha'}\delta M_{\text{sh}}}}{\delta M_{\text{sh}}}, 
\end{equation}
where $\delta M_{\text{sh}}$ is the mass of the shell as measured by a FIDO at the stretched horizon (at $\rho=\sqrt{\alpha'}$). For an observer at infinity, there is a redshift factor $\sqrt{\alpha'}/(4GM)$ and corresponding Hagedorn spectrum
\begin{equation}
\label{hagedos}
\rho(\delta M) \sim \frac{e^{\beta_{\text{Hawking}}\delta M}}{\delta M},
\end{equation}
with the Hagedorn temperature equal to the Hawking temperature. \\

\noindent From the Hagedorn density of states (\ref{hagedos}) with $\beta_H = \beta_{\text{Hawking}}$, the shell adds an entropy of
\begin{equation}
\delta S = \beta_{\text{Hawking}}\delta M
\end{equation}
which integrates to the Bekenstein entropy:\footnote{We focus on the Schwarzschild black hole here. Comments on the inclusion of $\alpha'$-corrections to the geometry are provided further on. The generalization of this argument to rotating black holes is provided in appendix \ref{rotat}.}
\begin{equation}
\delta S = 8\pi G M \delta M \to S = \frac{A}{4G}.
\end{equation}
This demonstrates that the long string is capable of yielding the correct number of states to account for the microstructure of the black hole. \\

\noindent Important early work on interpreting the black hole entropy in terms of fundamental strings is contained in \cite{Susskind:1993ws}\cite{Horowitz:1996nw}\cite{Larsen:1995ss}\cite{Maldacena:1996ya}\cite{Halyo:1996vi}\cite{Tseytlin:1996qg}. \\

\noindent The basic conceptual question is: how can an equilibrium for the shell around a black hole even exist in the first place? Is not everything perpetually falling inwards, never truly settling down to an equilibrium configuration? 
This is in principle so within field theory, as the thermal entropy of the quantum vacuum around the black hole diverges \cite{'tHooft:1984re}. Field modes Lorentz contract close to the horizon and allow an infinite amount of information to be stored. No sufficiently large radial pressure is built up in the process to compensate the increasing flattening of infalling matter. Not so for string theory. \\
The long string, as described by the thermal scalar, develops an energy density profile bound to the horizon (figure \ref{form}(a)), a feature impossible in QFT. The long string develops a pressure profile (figure \ref{form}(b)) and does not fall in further due to its failure to Lorentz contract close to the horizon \cite{Susskind:1993aa}. It therefore contributes a finite amount to the black hole entropy. \\

\noindent A second conceptual issue that we need to address is whether the already present thermal gas can influence the newly infalling gas in its equilibration process. We have seen in previous work \cite{Mertens:2013zya}\cite{Mertens:2014saa} that the most dominant contribution of the thermal atmosphere of the black hole comes from the string-length region very close to the horizon where the canonical gas at $T_{\text{Hawking}}$ percolates into long strings. It is widely accepted that the equilibration itself occurs on the stretched horizon, so our focus is on that part of the atmosphere.\\
Will the already present thermal gas not influence the energy and pressure profile of the new gas away from those displayed above in equation (\ref{stressen}) (where we did not consider the thermal atmosphere)? \\
The answer is no. To see this, we remind the reader of the seemingly innocent-looking property of the expression for the high-energy stress tensor derived above, that $T_{\mu\nu} \sim E$, with no $E$-dependence anywhere else. Now consider first equilibrating a (high-energy) gas with energy $E_1$ (in a generic space), and after it equilibrated we send in a second high-energy gas with energy $E_2$. The total energy density and pressure profile simply add with a resulting magnitude proportional to $E_1+E_2$.\footnote{Note that this is not a statement about extensivity as we are discussing the spatial profiles here.} This property is non-trivial, even for non-interacting matter. The reason is that in quantum statistical physics, Bose and Fermi statistics cause effective interactions that violate this property. As an illustrative situation, consider a gas of fermions with energy $E_1$ in a constant gravitational field. Adding a second gas will not simply give the same spatial profile, since zero-point pressure will force the second gas to stay away from locations where the first gas is localized. The resulting profile will hence be different than simply the algebraic sum. \\
The fact that this is not a property of the high-energy behavior of a string gas (in any spacetime) is closely related to the fact that a high-energy (string) gas obeys Boltzmann statistics. Another related point is that a highly excited string gas behaves as a random walk in space with $E\sim L$. Two random walks can combine together to form a random walk with total length simply the sum. \\
Coming back to the stretched horizon, we have here the same sort of mixing between two gases: the stretched horizon (being described by a canonical long string gas) and an additional microcanonical gas that we throw in. As shown above, their spatial profiles for energy and pressure will simply add up. The newly equilibrated gas has precisely the same energy and pressure profiles as shown in the previous section. \\ 
Physically, this means we throw long strings into the long string stretched horizon of the black hole. These additional long strings are simply absorbed by the already present long string(s). \\

\noindent Let us check the validity of our approximations. The energy of the microcanonical gas satisfies $\delta M \gg \beta_H^{-1}$ in the high-energy regime and with the Hagedorn temperature equaling the Hawking temperature, we need
\begin{equation}
\delta M \gg \frac{1}{GM}.
\end{equation}
Note that this is in fact a relatively small energy. But no matter how small the energy is measured at infinity, the blueshift will make the locally measured energy arbitrarily high and string-scale. \\
Also, we need this extra high-energy gas to be subdominant compared to the original black hole itself (small backreaction). This implies $M\gg \delta M$. In terms of the entropy, these inequalities are equivalent to
\begin{equation}
1 \ll \beta_H \delta M \ll \beta_{\text{Hawking}} M.
\end{equation}
This means the added entropy needs to be larger than 1, which means we need to add at least one degree of freedom. Secondly, the added entropy has to be much smaller than the black hole entropy. This obviously requires the original black hole entropy $\sim GM^2$ to be much larger than 1. Our argument hence applies to large black holes where the near-horizon Rindler approximation makes sense. \\

\noindent For large black holes with $\alpha'$-corrections, one can derive the Wald entropy under the plausible assumption that the zero-mode retains its marginal character for finite size black holes (i.e. not in the strict Rindler limit).\footnote{We are unable to prove this statement, although we would like to give two pieces of evidence for it. Firstly, the only (uncharged) black hole that is known sufficiently in string theory is the 2d $SL(2,\mathbb{R})/U(1)$ black hole. Here the zero-mode remains marginal for any size of the black hole. Secondly, in \cite{Giveon:2014hfa} arguments were given in favor of universality of this result, extending the marginality of the thermal scalar to a broader class of black hole solutions.} If one assumes this to be true, then it is straightforward to show that one can obtain the Wald entropy from this type of argument for the case where $\alpha'$-corrections to the geometry are included. String theory on the Euclidean geometry leads again to the conclusion that $T_H = T_{\text{Hawking}}$ where in this case the Hawking temperature of the higher-derivative black hole needs to be utilized. Since we are working on an infinitesimal level, we next find again the first law of black hole thermodynamics, which leads directly to the Wald entropy upon integration. In summary, this means that the previous derivation also works in these cases where one simply uses the Hawking temperature as determined by the $\alpha'$-exact black hole metric. \\

\noindent The above derivation correctly reproduces the Bekenstein-Hawking entropy. But is this not true for any matter falling in? After all, if one throws in some matter of energy $\delta E$ and entropy $\delta S$, then after a long time, the black hole's mass will have increased by $\delta E$ and its entropy by $\delta S$ where both quantities are related through the area law. This is basically the Zurek-Thorne argument where the infalling information that falls behind the stretched horizon is reinterpreted as the black hole entropy \cite{Zurek:1985gd}. \\
The main difference with our approach is that it is in general impossible to realize thermodynamical equilibrium of the infalling matter (with fixed energy and entropy) on its own. The first law of thermodynamics of the infalling matter, living in the heat bath of the black hole, states that
\begin{equation}
dE = T_{\text{Hawking}} dS,
\end{equation}
but on the other hand, the microcanonical temperature of the infalling gas is defined as
\begin{equation}
dE = T(E) dS,
\end{equation}
and hence, in general, $T(E) \neq T_{\text{Hawking}}$, so equilibrium is not realized immediately: the infalling matter either needs to deposit energy in the thermal bath or absorb energy from it before it can achieve equilibrium. \\

\noindent For long strings, we have $S \sim E$ (a Hagedorn density of states) and the microcanonical temperature is energy-independent. Hence, unless $T_H = T_{\text{Hawking}}$, it is \emph{impossible} to have thermodynamic equilibrium for the infalling long string gas. \\

\noindent The upshot is that only long strings (having a Hagedorn density of states) with $T_H = T_{\text{Hawking}}$ are automatically in equilibrium with the black hole already present, without interacting with it in a significant way. The long strings can hence be deposited on top of the existing black hole one by one, each one in equilibrium with the already present black hole. \\

\noindent Taking a step back, one can think of the black hole realized in this way as a set of equilibrated long strings. \\
Hence the black hole microstructure is realized in a quite natural fashion by these long strings. \\

\noindent To conclude this note, we would like to point out that our approach sheds some new light on the old puzzle that black holes appear to be described by $c=6$ CFTs \cite{Maldacena:1996ya}\cite{Halyo:1996vi}\cite{Tseytlin:1996qg}. It is interesting to note that the shift in the Hagedorn temperature from its flat space value $\beta_H = 2\sqrt{2}\pi\sqrt{\alpha'}$ to its value $\beta_H = 2\pi\sqrt{\alpha'}$ in Rindler space, as measured at the stretched horizon, precisely corresponds to the shift in central charge from $c=12$ to $c=6$. \\
\noindent Suppose in flat space, we inject a small amount of energy $\delta E \gg T_H$. This added gas is on its own a long string gas and hence adds an entropy of
\begin{equation}
\label{added1}
\delta S = 2\sqrt{2}\pi\sqrt{\alpha'}\delta E = 2\pi \sqrt{\frac{c}{6}}2\delta \sqrt{N},
\end{equation}
where the second equality is Cardy's formula. Equating then gives with $E = \frac{2}{\ell_s}\sqrt{N}$ that $c=12$, indeed the flat space type II central charge. \\

\noindent Next, suppose we add a small amount of string gas to a black hole, with energy $\delta E \gg T_{\text{Hawking}}$. Then the added entropy is of the form
\begin{equation}
\delta S = \beta_{\text{Hawking}} \delta E = 2\pi\sqrt{\alpha'}\delta E_{\text{sh}} = 2\pi \delta E_{R},
\end{equation}
where $E$ is the energy measured at infinity, $E_{\text{sh}}$ is the energy measured at the stretched horizon and $E_R$ is the dimensionless energy measured at $\rho=1$. The black hole itself can be interpreted as a long string in flat space, where the Rindler energy $E_R$ and the oscillation number $N$ of this long string (not the worldsheet CFT of the Rindler string!) are related as $E_R = 2\sqrt{N}$, such that $E_{\text{sh}} = \frac{2}{\ell_s}\sqrt{N}$ \cite{Maldacena:1996ya}\cite{Halyo:1996vi}. Hence the added entropy is related to an added oscillation number of the long string making up the black hole. It can be written as
\begin{equation}
\delta S = 4\pi \delta \sqrt{N} = 2\pi \sqrt{\frac{c}{6}} 2 \delta \sqrt{N},
\end{equation}
immediately leading to $c=6$. \\
The different Hagedorn temperatures for flat space versus Rindler space are what ultimately cause the $c=12 \to c=6$ shift of the central charge. \\
\emph{The universality of $c=6$ is hence directly related to the universality of $T_H = T_{\text{Hawking}}$.}

\section{Some comments and conclusion}
\label{concl}
The arguments in this paper are very general and do not depend on the details of the black hole solution. The only condition that has to be fulfilled is that the metric is Rindler close to the horizon. In this case, as we showed for type II and heterotic strings in \cite{Mertens:2013zya}\cite{Mertens:2014saa}, the single string has a Hagedorn asymptotic density of states with $T_H = T_{\text{Hawking}}$ because the thermal scalar has a zero-mode at this temperature. The situation is more intricate for bosonic strings and we avoid making any statement about them at this point \cite{Mertens:2013zya}. It is reasonable to assume that the existence of the zero-mode for the thermal scalar in Rindler space reflects the stability of the Minkowski vacuum. \\

\noindent The long string shell we consider is not self-interacting. The only interaction we take into account is between the shell and the background metric. In a sense, this is a mean-field approximation where the classical Einstein equations are used to calculate the mean-field background metric and the single string Schr\"odinger equation in the background determines the single string density of states and hence the number of different ways in which a small amount of mass can be added to the black hole. 
Furthermore, we needed a non-zero but small string interaction $g_s$ to let the shell equilibrate to a long string. How does inclusion of interactions affect the classical black hole entropy? In \cite{Susskind:1994sm}, Susskind and Uglum discussed the renormalization of the classical black hole entropy to all orders in the string coupling constant in the canonical formalism. They argued that small loops which  encircle  the conical singularity introduced by varying $\beta$ and contribute to the entropy, renormalize Newton's constant. We expect that an analogous reasoning using the microcanonical formalism can show that the contribution of residual interactions beyond mean field to the entropy, will likewise renormalize Newton's constant. \\

\noindent It seems clear that the equality of the Hawking temperature and Hagedorn temperature is not accidental, but encodes some important features on how long strings behave in gravitational fields and on how long strings can give the necessary microstructure to black holes. It will be interesting to investigate this further.

\section*{Acknowledgements}
We thank K. Van Acoleyen for some interesting suggestions and discussions. TM acknowledges financial support from the UGent Special Research Fund, Princeton University, the Fulbright program and a Fellowship of the Belgian American Educational Foundation. The work of VIZ was partially supported by the RFBR grant 14-02-01185.

\appendix
\section{Rotating black holes}
\label{rotat}
\subsection{Preliminaries: geometry and thermal ensemble}
As noted before, Rindler space is the near-horizon region of a general uncharged black hole. It is also the near-horizon region of a rotating black hole (e.g. a Kerr black hole) in terms of the co-rotating ZAMO observers. Without going into details, the Kerr metric is of the form (see e.g. \cite{Misner:1974qy})
\begin{equation}
ds^2 = - \alpha^2 dt^2 + \frac{\rho^2}{\Delta}dr^2 + \rho^2 d\theta^2 + \bar{\omega}^2\left(d\phi - \omega dt\right)^2,
\end{equation}
where $\alpha$, $\rho$, $\Delta$, $\omega$ and $\bar{\omega}$ are functions of $r$ and $\theta$ whose precise form will not interest us. Close to the horizon, one can approximate $\omega \approx \Omega_H$, and define a new angular coordinate as $\bar{\phi} = \phi - \Omega_H t$. Then a redefinition of the radial coordinate directly leads to Rindler spacetime. The main point is then that in the original coordinates, the near-horizon Rindler region is moving along with the horizon angular velocity $\Omega_H$ in the $\phi$-direction. Just as before, the Kerr metric itself is not a solution of string theory but Rindler space is. \\

\noindent Let us now remind the reader of how one would define a thermal ensemble for such black holes. \\
The timelike Killing vector $\frac{\partial}{\partial t}$ turns spacelike within the ergosphere. For a rotating black hole, the horizon is instead a Killing horizon of the co-rotating vector field $\chi = \frac{\partial}{\partial t} + \Omega_H \frac{\partial}{\partial \phi}$ which does remain timelike inside the ergoregion. It is this vector field that is associated with the Hartle-Hawking vacuum (supposing it can be sensibly defined). This vector field, however, becomes spacelike far away (unlike $\frac{\partial}{\partial t}$), corresponding to the fact that rigid rotation automatically implies faster-than-light travel for distant observers. For a hole that rotates sufficiently slow, this region is arbitrarily far away and a sensible quantization might be allowed \cite{Frolov:1989jh}. This does raise some questions regarding the level of rigor with which the Hartle-Hawking vacuum can actually be constructed. On the other hand, in AdS backgrounds, rotating black holes exist for which the co-rotating Killing vector field remains timelike throughout the full exterior of the hole \cite{Hawking:1999dp}\cite{Winstanley:2001nx}. \\

\noindent In any case, suppose we start not with a canonical ensemble but with a grand canonical ensemble of the gas around the black hole where the angular momentum of the mode contributes to the partition function as:
\begin{equation}
\text{Tr}e^{-\beta\left(H - \Omega_H J\right)}.
\end{equation}
Then this can be viewed equivalently as a canonical ensemble using $\chi$ as the new time translation operator:
\begin{equation}
\text{Tr}e^{-\beta\left(H - \Omega_H J\right)} = \text{Tr}e^{-\beta H'},
\end{equation}
where we have denoted the new Hamiltonian by $H'$. Note that for this interpretation to be possible, it is vital that we start with a grand canonical ensemble with angular velocity equal to the black hole angular velocity $\Omega_H$. In somewhat more detail, suppose we have a mode in the spectrum with energy $\omega$ and angular momentum $m$:
\begin{equation}
\psi_{\omega,m} \sim e^{-i\omega t}e^{im\phi}.
\end{equation}
This contributes to the grand canonical trace as $e^{-\beta\left(\omega - \Omega_H m\right)}$. Performing the coordinate redefinition $\bar{\phi} = \phi - \Omega_H t$, the mode can be written as
\begin{equation}
\psi_{\omega,m} \sim e^{-i(\omega -m\Omega_H) t}e^{im\bar{\phi}},
\end{equation}
and it now satisfies $e^{-\beta H} = e^{-\beta\left(\omega - \Omega_H m\right)}$. Hence its contribution equals that of a canonical ensemble after the coordinate redefinition. Also, one readily finds that the above coordinate transformation indeed yields
\begin{equation}
\chi = \frac{\partial}{\partial t} + \Omega_H \frac{\partial}{\partial \phi} \,\, \to \,\, \frac{\partial}{\partial t}.
\end{equation}

\subsection{The rotating long string}
The remaining steps to discuss the thermal scalar and the random walk in this case are now straightforward, although, due to the above reservations, they are not airtight. \\
For a large enough black hole, or anticipating that the near-horizon region will contain the most dominant contribution, we can approximate the problem again by zooming in to the near-horizon region. We can make the transition to the near-horizon co-rotating ZAMOs and their mode spectrum, quantized by $\chi$.\footnote{$\chi$ coincides with ZAMO movement close to the horizon. When going further away, $\chi$ describes a rigid rotation whereas the ZAMO's rotation slows down (according to an asymptotic observer) as their worldlines move further out. At infinity, a ZAMO is not rotating at all anymore (w.r.t. the coordinate $\phi$).} These observers effectively see a static Rindler region. \\ 

\begin{figure}[h]
\centering
\includegraphics[width=0.5\textwidth]{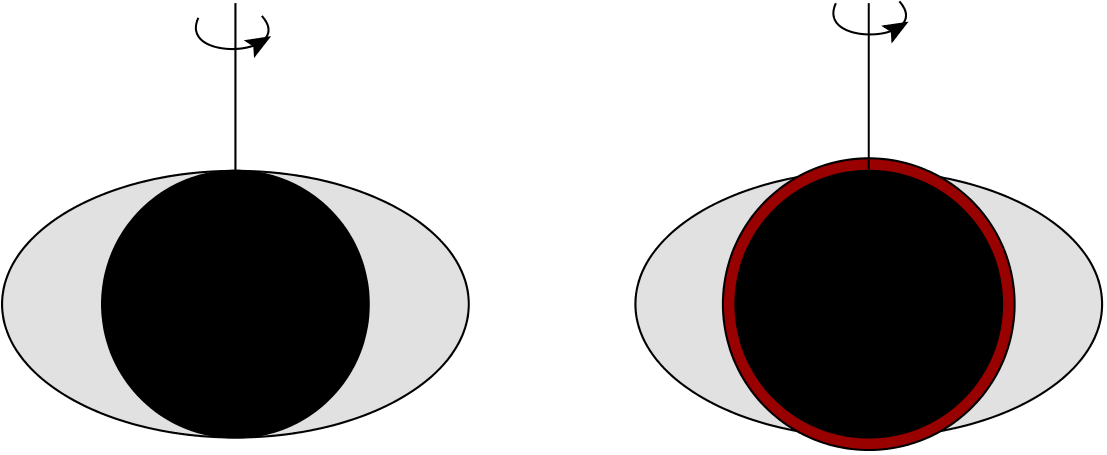}
\caption{Left figure: structure of a rotating black hole. $\frac{\partial}{\partial t}$ becomes spacelike in the ergoregion (the gray region), whereas $\chi$ remains timelike in that region. Right figure: most dominant contribution to the grand canonical trace. The Rindler thermal scalar provides the dominant contribution (colored in red) and rotates rigidly along with the hole itself.}
\label{rota}
\end{figure}

\noindent In any case, the canonical trace is dominated by the thermal scalar in the near-horizon Rindler region. Hence the full grand canonical trace, as constructed at infinity by non-moving observers (or ZAMOs at infinity), is dominated by the thermal scalar mode. As this mode has no non-trivial profile tangential to the horizon, its wavefunction is unaltered as the hole rotates, and the long string again forms a region close to the horizon of string-scale thickness. Thus from infinity, the string-sized shell surrounding the black hole simply rotates rigidly along with the hole itself. The full situation is sketched in figure \ref{rota}.

\noindent Let us finally show how the first law is encoded in this formalism. We have found that 
\begin{equation}
\text{Tr}e^{-\beta (H -\Omega_H J)} = \text{Tr}e^{-\beta H'} \approx Z_{\text{th.sc.}},
\end{equation}
where the grand canonical ensemble is transformed into a canonical ensemble in a static spacetime, which in turn is dominated by the thermal scalar contribution close to the horizon. Inverse Laplace transforming then yields\footnote{The Jacobian in going from $(E',J)$ to $(E,J)$ is trivial.}
\begin{equation}
\rho(E,J) \sim e^{\beta_{\text{Hawking}}E'} \sim e^{\beta_{\text{Hawking}}(E-\Omega_H J)}.
\end{equation}
So the density of states of a string gas in a rotating black hole geometry (with $\beta_{\text{Hawking}}$ and $\Omega_H$) at high energies $E'$ or $E \gg \Omega_H J$ is of the above form. Hence throwing in some long stringy matter with energy $\delta E$ and angular momentum $\delta J$ into the pre-existing rotating black hole leads to an increment of the thermal entropy of
\begin{equation}
\delta S = \beta_{\text{Hawking}} \left(\delta E - \Omega_H \delta J\right).
\end{equation}
Slightly rearranging this, one indeed finds
\begin{equation}
\delta E = T_{\text{Hawking}} \delta S + \Omega_H \delta J,
\end{equation}
which is again the first law of black hole thermodynamics. Hence, once again, the equilibrated long strings can account for the black hole entropy.

\end{document}